\documentclass[9pt,twocolumn,twoside]{opticajnl}
\journal{opticajournal} 

\setboolean{shortarticle}{false}

\usepackage{lineno}
\usepackage{graphicx}
\usepackage{subfigure}
\usepackage{appendix}
\usepackage{comment}
\usepackage{caption}
\usepackage{subcaption}
\usepackage{overpic}

\title{Long distance local local oscillator continuous variable quantum key distribution  with digital signal processing}

\author[1,2]{Dengke Qi}
\author[1,*]{Xiangyu Wang}
\author[1]{Jiayu Ma}
\author[1]{Zhenghua Li}
\author[3]{Ziyang Chen}
\author[2]{Yueming Lu}
\author[1]{Song Yu}

\affil[1]{State Key Laboratory of Information Photonics and Optical Communications, Beijing University of Posts and Telecommunications, Beijing 100876, China}
\affil[2]{School of Cyberspace Security, Beijing University of Posts and Telecommunications, Beijing 100876, China}
\affil[3]{State Key Laboratory of Advanced Optical Communication Systems and Networks, School of Electronics, and Center for Quantum Information Technology, Peking University, Beijing 100871, China}

\affil[*]{xywang@bupt.edu.cn}

\begin{abstract}
Quantum key distribution relying on the principles of quantum mechanics enables two parties to produce a shared random secret key, thereby ensuring the security of data transmission. Continuous variable quantum key distribution (CV-QKD) is widely applied because it can be well combined with standard telecommunication technology. Compared to CV-QKD with a transmitting local oscillator, the proposal of CV-QKD with a local local oscillator overcomes the limitation that local oscillator will attenuate as transmission distance increases, providing new possibilities in long-distance transmission. However, challenges still persist in practical long-distance transmission, including data sampling and recovery under low signal-to-noise ratio conditions. In order to better recover data and reduce the additional excess noise, we propose the least squares fitting algorithm to get more accurate sampling data and complete more accurate phase compensation.
Herein, we demonstrate the long-distance local local oscillator CV-QKD experiment which have considered the effect of finite-size block over 120 km of standard optical fiber  with high efficient real-time post-processing. The results not only verify the good performance of the system over long distance, but also paves the way for large-scale quantum secure communications in the future.
\end{abstract}

\setboolean{displaycopyright}{false} 

\begin{document}

\maketitle

\section{Introduction}
Quantum key distribution (QKD) enables two legitimate communicating parties to share a secure key over an insecure quantum channel, with the unconditional security of the key guaranteed by the laws of quantum mechanics. Currently, there are two methods for key distribution: discrete variable QKD and continuous variable QKD (CV-QKD) \cite{grosshans2002continuous}. As a branch of QKD, CV-QKD has garnered increasing attention in the field of quantum science because it can easily integrate with classical coherent optical communication, such as signal modulation and detection. In recent years, significant progress has been made in both theoretical research \cite{grosshans2003quantum,weedbrook2004quantum,leverrier2009unconditional,leverrier2010finite,furrer2012continuous,leverrier2015composable,leverrier2017security,lin2019asymptotic,pirandola2021composable,wang2023non} and experimental implementations \cite{jouguet2013experimental,qi2015generating,huang2016long,kleis2017continuous,laudenbach2019pilot,ren2021demonstration,tian2022experimental,pan2022experimental,wang2023experimental,Xu:23,liu2023experimental,hajomer2024long,qi2024experimental} of CV-QKD. Especially, a new digital signal processing analysis method has recently been proposed, allowing CV-QKD to be better integrated with classical coherent optical communication \cite{chen2023continuous}.

Since the inception of CV-QKD, quantum signals and local oscillator (LO) have been transmitted together which is called the transmitting local oscillator (TLO) CV-QKD for extended periods, ensuring stable interference and lower phase noise. However, this approach is vulnerable to attacks from Eve, such as LO fluctuations \cite{ma2013local} or calibration attacks \cite{jouguet2013preventing}. Moreover, the LO also attenuates along with the quantum signals during transmission. Especially, LO have insufficient power to meet the limit of shot-noise-limited detection of the quantum signals over long distances. Until 2015 \cite{qi2015generating,soh2015self}, a new scheme with a simpler structure and capable of avoiding the aforementioned security vulnerabilities was proposed for the first time. In this scheme, LO and the signal are generated by two separate lasers. Hence, it is referred to as local local oscillator (LLO) CV-QKD. While LLO CV-QKD can overcome the drawbacks of TLO CV-QKD, it must also face the challenge of eliminating frequency offsets between the LO and quantum signals. Since the first two proposals, various pilot schemes are designed to reduce the frequency offset and phase noise, such as pilot-sequential and pilot-multiplexing LLO CV-QKD and numerous LLO CV-QKD experiments \cite{laudenbach2019pilot,pan2022experimental,jain2022practical,hajomer2024long} have been realized. Especially, recent experiment \cite{hajomer2024long} has claimed they have successfully demonstrated LLO CV-QKD over 100 km of ultralow-loss optical fiber by controlling the phase noise–induced excess noise through a machine learning framework for carrier recovery and optimizing the modulation variance. However, as the transmission distance increases, the final data processing and recovery become more challenging. On one hand, it's difficult to maintain stable time synchronization between Alice and Bob over long distances. Clock jitter and drift between them inevitably occur due to external factors, leading to sampling data deviations. On the other hand, the precision of phase compensation decreases due to low signal-to-noise ratios, making accurate data recovery challenging. In this case, it leads to the introduction of more excess noise into the entire system. However, compared to short distances, lower levels of excess noise are more critical for long-distance scenarios. Therefore, reducing the excess noise to get good performance is also crucial for LLO CV-QKD in long-distance scenarios.

Here in our work, we separately demonstrate the LLO CV-QKD experiment over a standard optical fiber of 80 km and 120 km. Considering the impact of low signal-to-noise ratio at the current distance, clock jitter in the time synchronization between both sides and the need for lower excess noise, we conducted an excess noise analysis to better control it and enhance the performance of the system as much as possible. Hence, we proposed a method for data recovery: the least squares algorithm \cite{dyer2001least}. This approach enabled Bob to obtain data closer to the real values, facilitating more precise phase compensation and reducing the extra introduced excess noise. In addition, we designed the high efficient post-processing procedure to generate the key rate more efficiently in real-time and achieve  sufficiently high secret key rates at long transmission distances. Based on this, we achieved excess noise at the milli shot noise level and a secret key rate of 76 kbps with the block size of $1\times10^9$ over 80 km. In addition, we could increase the actual transmission distance and finally successfully achieve a secret key rate of 5.36 kbps with the block size of $1\times10^{11}$ over 120 km. 

Our work provides a feasible way of recovering accurate data under low signal-to-noise ratio. It can also break further distances and pave the way for large-scale quantum secure communications in the future, thus reinforcing the importance of our work.

\section{Experiment}

\subsection{Experiment Setup}
\begin{figure*}[h]
\centering
{\includegraphics[width=\linewidth]{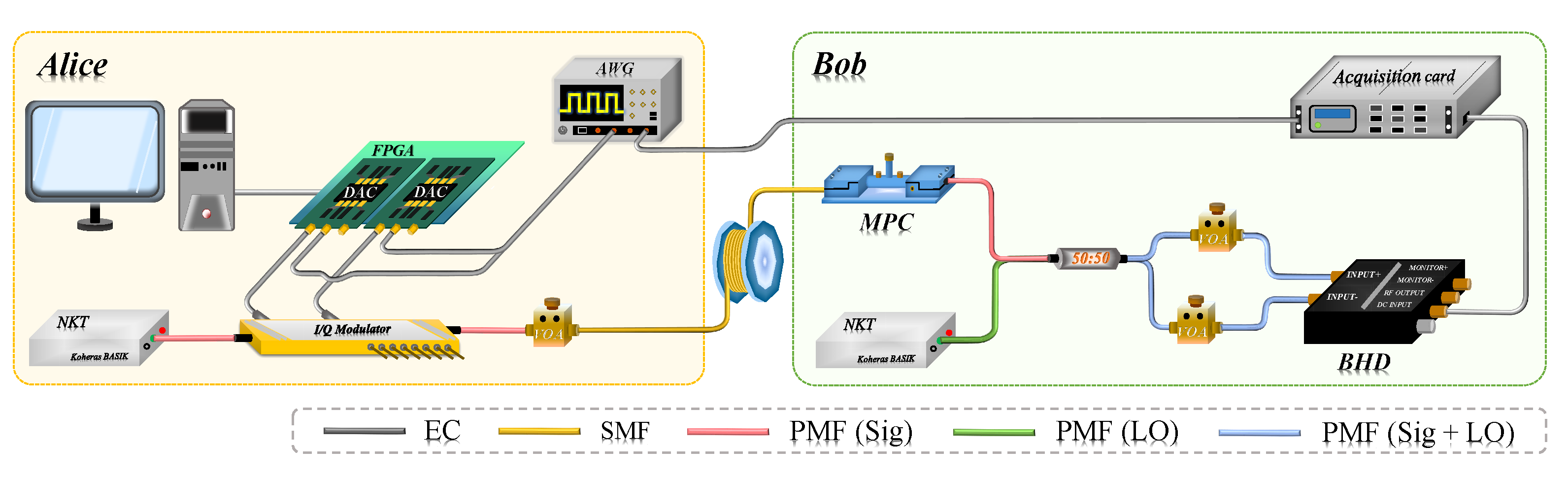}}
\caption{The optical layout of GMCS LLO CV-QKD experimental scheme. DAC: digital-to-analog converter, IQ modulator: in phase/quadrature modulator, AWG: arbitrary waveform generator, MPC: manual polarization controller, VOA: optical variable attenuator, BHD: balanced homodyne detector.}
\label{fig:experiment}
\end{figure*}

The optical layout of the Gaussion modulation coherent sate (GMCS) LLO CV-QKD experiment is shown in Fig. \ref{fig:experiment}. At the Alice's side, a continuous wave laser whose wavelength is 1550 $nm$ with a narrow linewidth is initially employed to generate coherent light. Subsequently, the coherent light enters an in-phase/quadrature (IQ) modulator to generate a GMCS with a repetition frequency of 100 MHz. To better compensate for fast phase drift, the signal here includes pilot tone and quantum signals with a ratio of 1:1. So, the actual repetition frequency is 50 MHz. The modulation signal of the IQ modulator is generated by a self-developed 16-bit DAC board mounted on an FPGA. It is worth noting that we adopt pilot-sequential LLO structure, hence the signals herein comprises both quantum signal and pilot signal. To enhance the quality of the signals, we utilize a bias controller to dynamically stabilize the modulator over a long period, acquiring the optimal modulated signals. Next, a variable optical attenuator (VOA) was placed after the IQ modulator to adjust the modulation variance of the signals. Finally, the signals were sent through a quantum channel consisting of standard optical fibers, where the length of the fiber is 80 km and 120 km with an average attenuation of 0.2 dB/km. Among them, the modulation variance at 80 km is 4.01 SNU and the modulation variance at 120 km is 9.41 SNU.

At the Bob's side, in order to obtain a high-quality signal to reduce noise as much as possible, we placed an manual polarization controller (MPC) to align the state of polarization (SOP) of the quantum signal. A continuous wave laser which is the same as the Alice's side was used to generate LO. The power of LO in our experiment was large enough to meet the limit of shot-noise-limited detection of the signals. Next, the LO and the signals from Alice were combined on a beam splitter (BS). Then, the output signals from the BS are passed through identical lengths of VOA to promise the automatic balance of the two arms, before being sent into a balanced homodyne detector (BHD) with a bandwidth of 1.6 GHz. Finally, the output signal from the BHD was acquired by a continuous acquisition card operating at a sampling rate of 1 Gsa/s. In our experiment, FPGA and acquisition card were triggered by the same synchronous source to achieve clock synchronization at the transmitter and receiver for simplicity. In our experiment, FPGA and acquisition card were synchronized using a 10-MHz clock which is provided by AWG. It was worth noting that in order to consider finite-size effects and make subsequent data processing work more convenient, we deliberately chose the acquisition card which can stably acquire data continuously for a long time.
\subsection{Digital Signal Processing}

\begin{figure}[h]
\centering
{\includegraphics[width=\linewidth]{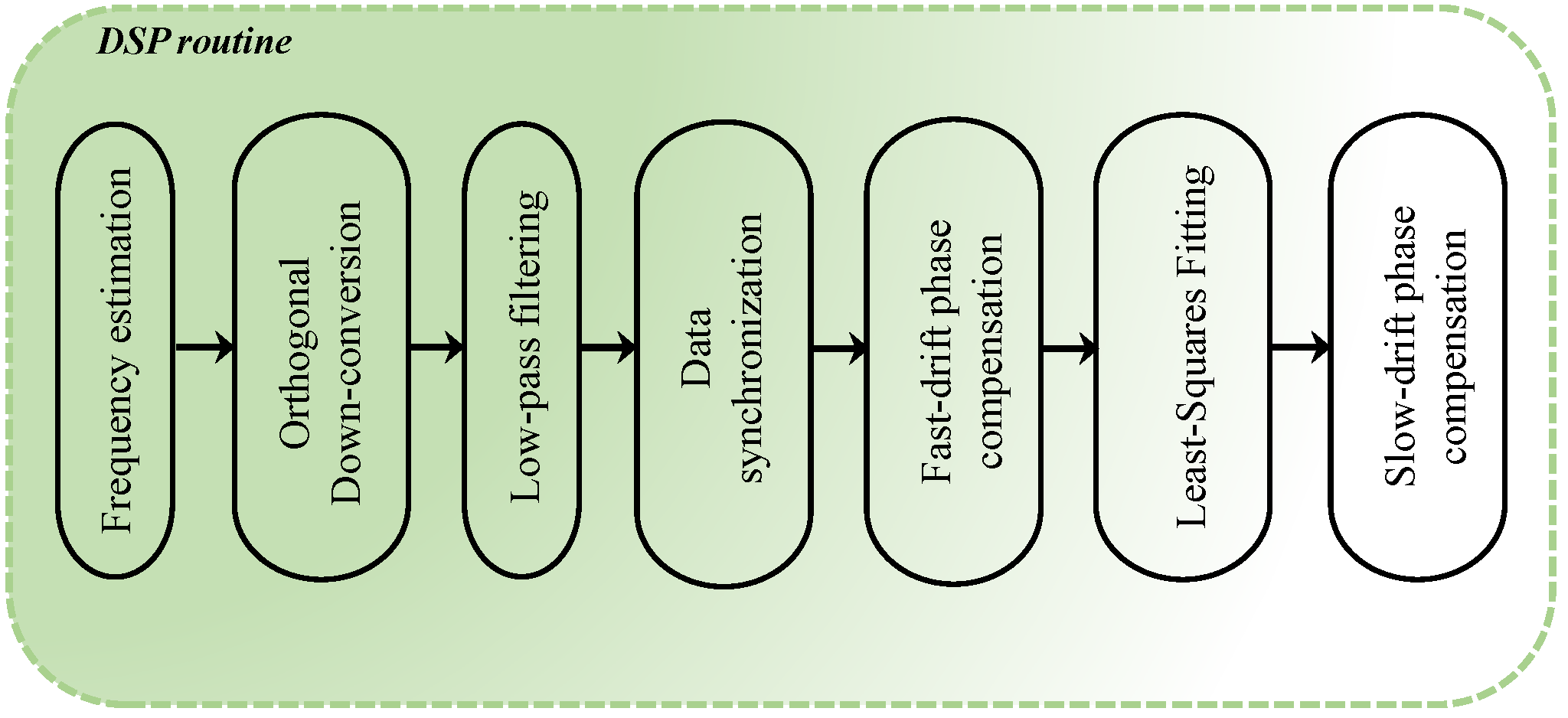}}
\caption{DSP routines of LLO CV-QKD through least squares fitting assisted data recovery over 80 km and 120 km.}
\label{fig:dsp}
\end{figure}
\begin{figure}[h]
\centering
{\includegraphics[width=0.75\linewidth]{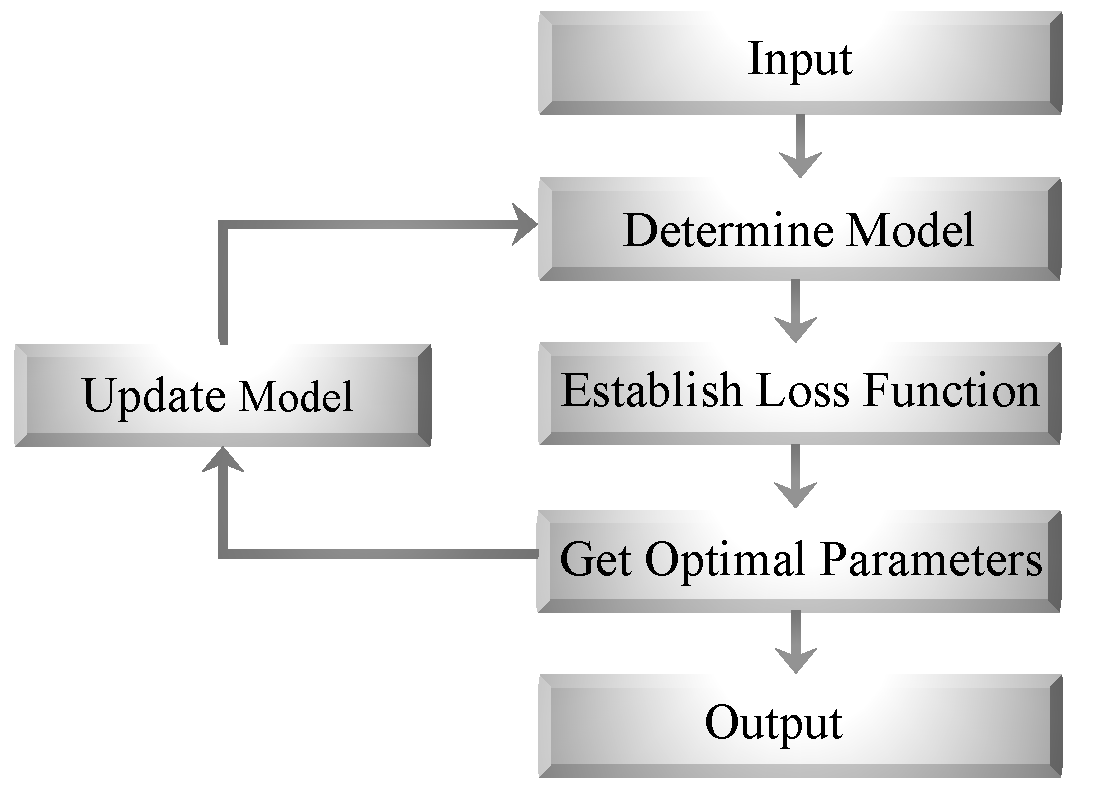}}
\caption{Process of least square.}
\label{fig:LS}
\end{figure}

Fig. \ref{fig:dsp} shows DSP routines of LLO CV-QKD through least squares fitting assisted data recovery (the detailed derivation process can be found in \ref{app:a}.). Firstly, we performed a Fast Fourier Transform  on the signal to determine the frequency offset between the LO and the signal based on their spectra. Subsequently, using the obtained frequency offset which is around 300 MHz, we down-convert the quantum signal and the pilot tone separately. Then, we chose an appropriate bandwidth based on the spectrum range of the obtained signals for the low-pass filter to obtain the corresponding X and P orthogonal components of the quantum signal and the pilot tone $X_\textup{sig}$ and $P_\textup{sig}$. Next, we proceed with phase compensation for the fast phase drift caused by frequency offset. As shown in Fig. \ref{fig:dsp}, before performing slow drift phase compensation, we apply a least squares method to compensate for sampling deviations of the signals caused by external clock jitter, thereby facilitating more effective slow drift phase compensation in the subsequent steps. The process of least square\cite{dyer2001least} is shown in Fig. \ref{fig:LS}. First, we take $X_\textup{sig}$ and $P_\textup{sig}$  as inputs to the entire model. Given that the signal preparation frequency is 100 MHz and the sampling rate is 1 GSa/s, each pulse signal yields 10 sampling points.So, we denote each pulse sequence as \(\left\{x_1,x_2,\cdots x_9,x_{10}\right\}\) and \(\left\{p_1,p_2,\cdots p_9,p_{10}\right\}\) with corresponding instants denoted as \(\left\{t_1,t_2,\cdots t_9,t_{10}\right\}\). Specifically, we use the least squares method to fit the sample points within each pulse. Taking $X_\textup{sig}$ as an example, for each pulse, we first should determine the curve model, here we start with a third-order polynomial fit. So, the curve can be expressed as $\widehat{x_i}=at_i^3+bt_i^2+ct_i+d$, which needs to satisfy:
\begin{equation}
\centering
\hat{a},\hat{b},\hat{c},\hat{d}=\textup{argmin}_{a,b,c,d} \left(\sum_{i=1}^{n}\left(x_{i}-\hat{x_{i}}\right)^{2}\right)
\end{equation}
Where \(a\), \(b\), \(c\), and \(d\) are the fitting parameters, the function $\textup{argmin}$ represents finding the minimum argument of the corresponding independent variable.
To measure the quality of the model, a loss function which is also the core concept of the least squares method is typically used. The loss function of the least squares method is typically defined as the sum of the squared errors between the predicted and actual values, can be expressed as 
\begin{equation}
\centering
Q=\sum_{i=1}^{n} \left(x_i-at_i^3-bt_i^2-ct_i-d\right)^2
\end{equation}
Next, we find the optimal parameters $a,b,c,d$ that minimize $Q$ in the current model. Therefore, we take partial derivatives with respect to each of these parameters as follows
\begin{equation}
\left\{\begin{array}{l}
\frac{\partial Q}{\partial a}=-2 \sum_{i=1}^{n} t_{i}^{3}\left(x_{i}-a t_{i}^{3}-b t_{i}^{2}-c t_{i}-d\right)=0 \\
\frac{\partial Q}{\partial b}=-2 \sum_{i=1}^{n} t_{i}^{2}\left(x_{i}-a t_{i}^{3}-b t_{i}^{2}-c t_{i}-d\right)=0 \\
\frac{\partial Q}{\partial c}=-2 \sum_{i=1}^{n} t_{i}\left(x_{i}-a t_{i}^{3}-b t_{i}^{2}-c t_{i}-d\right)=0 \\
\frac{\partial Q}{\partial d}=-2 \sum_{i=1}^{n}\left(x_{i}-a t_{i}^{3}-b t_{i}^{2}-c t_{i}-d\right)=0
\end{array}\right.
\end{equation}
Exactly, we can obtain the values of $a,b,c$ and $d$ that minimize the loss function. These optimal values correspond to the optimal parameters $\hat{a},\hat{b},\hat{c}$ and $\hat{d}$. Finally, the fitting curve can be obtained by bringing the corresponding parameters into the fitting curve model. After attempting third-order fitting, we also tested higher-order fits and repeated the above steps multiple times. However, we found that third-order fitting was sufficient to improve the results, and further increasing the fitting order could lead to overfitting. Therefore, in order to ensure fitting accuracy while minimizing computational complexity and processing time, we finally chose third-order fitting. For other nonlinear fitting methods, such as Fourier fitting or exponential fitting, the complexity is relatively high. Polynomial fitting can significantly reduce computational consumption and allow for real-time processing. Therefore, in our work, we ultimately adopted third-order polynomial fitting. Similarly, the same process is completed for $P_\textup{sig}$.

After LS, we use a phase searching algorithm to compensate slow drift phase (the details can be found in \ref{app:a}.). 

\subsection{ Real-time High-efficient Post-processing}

In order to generate key rate more efficiently in real-time, we designed the following high efficient post-processing procedure which is also needed to achieve sufficiently high secret key rates at long transmission distances. After quantum state detection, the data is classical data, and the core steps of post-processing are information reconciliation and privacy amplification, which can be realized based on a block-processing approach \cite{wang2018high,wang2018high1}. This means that it is not necessary to store enough data to start the post-processing process, and it can be processed while collecting, and finally complete data processing through an equivalent method in mathematics. After both communicating parties obtain a set of correlated raw keys, in order to reduce performance loss under low signal-to-noise ratio, multidimensional reconciliation is employed to convert continuous Gaussian variables into binary data. The nonzero elements of the sparse matrix involved in the computation are stored as one-dimensional arrays corresponding to their respective ordinate coordinates, thereby reducing the computational complexity from $O(n^3)$ to $O(n^2)$. After completing reconciliation, QC-LDPC \cite{milicevic2018quasi} code is applied to error-correct the binary sequence to obtain a consistent key. By using a layered belief propagation algorithm, the iteration count is reduced to half of the global decoding algorithm, and parallel computation of codewords can be implemented on a graphic processing unit (GPU). It is worth mentioning that before error-correction, we reordered the binary sequence to allow continuous threads in the GPU to read and write data in continuous memory space, thus improving addressing efficiency. We also use half-precision data types to reduce data storage and circulation overhead, greatly improving decoding throughput with slight loss of precision. Finally, we achieve the decoding speed to 67.13 Mpps (pulse per second). It is clear that the decoding speed we have obtained can fully support our experimental system. After error-correction, the data no longer compare the syndromes but directly proceeds to privacy amplification. Specifically, in the step of privacy amplification, we use hash function (Toeplitz matrices in our scheme) to distill the final key at speed 1.35 Gbps using GPU. In this way, the time of calculating the syndromes and communication between Alice and Bob are greatly reduced, and the final secret key rate is not affected. Our high efficient post-processing is implemented based on asynchronous architecture of central processing unit and GPU. Moreover, the calculation of decoding initial value and the syndromes in the procedure are also implemented through GPU. Through the proposed method, the time of post-processing can be reduced to more than half, greatly improving the overall efficiency which supports high-speed real-time CV-QKD under low signal-to-noise ratio.

\section{Noise Analysis And Results}
\subsection{Excess Noise Analysis}

The magnitude of excess noise directly affects the final performance, especially at a long distance, so excess noise analysis is crucial. Excess noise in LLO CV-QKD can arise from various sources, including quantization, modulation, relative intensity noise, phase noise and sampling-error noise. Assuming these noise sources are statistically independent, therefore, the total excess noise can be expressed as the sum of individual contributions:
\begin{equation} 
\begin{aligned}
\varepsilon_\textup{total}=\varepsilon_\textup{RIN}+\varepsilon_\textup{DAC}+\varepsilon_\textup{Mod}+\varepsilon_\textup{Phase}+\varepsilon_\textup{sam-error}
\end{aligned}
 \end{equation}
where $\varepsilon_\textup{RIN}$ is the intensity noise of two independent lasers, $\varepsilon_\textup{DAC}$ is digital-to-analog conversion (DAC) quantization noise, $\varepsilon_\textup{Mod}$ is the  modulation noise, $\varepsilon_\textup{Phase}$ is the phase noise and $\varepsilon_\textup{sam-error}$ is the noise caused by the error of the sampling signals.

The first term $\varepsilon_\textup{RIN}$ represents the intensity noise originating from the two independent lasers. The $\varepsilon_\textup{RIN}$ of a laser describes random fluctuations in the output optical power of the laser. Specifically, it is the ratio of the fluctuations in the output optical power to the average optical power. This noise represents the change in the output power of the laser over time and is usually related to the internal noise source of the laser. It can be expressed as \cite{laudenbach2018continuous}
\begin{equation}
 \begin{aligned}
\varepsilon_\textup{RIN}&=\varepsilon_\textup{RIN,sig}+\varepsilon_\textup{RIN,LO} 
\\
    &=T V_\textup{A} \sqrt{\textup{RIN}_\textup{sig} B_\textup{sig}}+\frac{1}{4}\textup{RIN}_\textup{LO} B_\textup{LO} V_\textup{$\neg$RIN,LO}(\hat{q})
 \end{aligned}   
\end{equation}

where $T$ is the total channel transmittance, $\textup{RIN}_\textup{sig}$ and $\textup{RIN}_\textup{LO}$ are the relative intensity noise of Alice's and Bob's laser, respectively, $B_\textup{sig}$ and $B_\textup{LO}$ correspond to their optical bandwidth of their lasers. Moreover, $V_\textup{$\neg$ RIN,LO}(\hat{q})=TV_\textup{A}$ represents the variance of $\hat{q}$ without taking count of the LO's RIN. In our work, the lasers both have a low RIN of -135dBc@10 MHz and the optical bandwidth is around 0.1 KHz. So, $\varepsilon_\textup{RIN}$ can be calculated to be around $1.79\times10^{-7}$ SNU and $0.66\times10^{-7}$ SNU at 80 km and 120 km ,respectively.

The second term $\varepsilon_\textup{DAC}$ is the DAC quantization noise. The noise is introduced when a digital signal is converted to an analog signal by a DAC. This noise stems from the discreteness of the DAC, which converts a continuous digital signal into a discrete analog signal, resulting in quantization error. Quantization error occurs because the output of the DAC is of limited precision and cannot accurately represent the continuous input signal. This limitation means that the DAC cannot precisely represent the amplitude of certain input digital signals, thus generating quantization noise. Since when the digital signal generated by the FPGA passes through the DAC in our work, due to the limitation of the quantization bits, there is a certain error between the analog modulated signal and the ideal signal, given by
\begin{equation}
   \varepsilon_\textup{DAC}=V_\textup{A}\left[\pi\frac{\delta V_\textup{DAC}}{V_\textup{DAC}}+\frac{\pi^{2}}{2}\left((\frac{\delta V_\textup{DAC}}{V_\textup{DAC}}\right)^{2}\right]^{2}      
\end{equation}
where $V_\textup{DAC}$ is the output voltage of a self-developed DAC board mounted on an FPGA, $\delta V_\textup{DAC}$ represents the voltage error due to the limited quantization bits. In our work, the DAC board is developed by ourselves which has high quantization bits naturally resulting in low voltage error. So, $\varepsilon_\textup{DAC}$ can be calculated to be around $3.56\times10^{-6}$ SNU and $8.37\times10^{-6}$ SNU at 80 km and 120 km ,respectively.

The third term $\varepsilon_\textup{Mod}$ is modulation noise. In the pilot-sequential LLO CV-QKD, the strong pilot tone and the weak quantum signal are generated by the same modulator. However, in practice, the limited dynamics of the modulator will result in an additional noise to the signal. Specifically, when we generate pilot tone and quantum signal, due to the finite dynamics of the modulator, if there is not enough extinction ratio, it will introduce leakage to the weak quantum signal, thus quantitates the noise as \cite{marie2017self}
\begin{equation}
  \varepsilon_\textup{Mod}=\left|a_\textup{s}\right|^{2}10^{-d_\textup{dB}/10}
\end{equation}

where $a_\textup{s}$ is the amplitude of the quantum signals and where $d_\textup{dB}$ represents the extinction ratio of the in phase/quadrature (IQ) modulator, can be calculated as
\begin{equation}
    d_\textup{dB}=10\textup{log}_\textup{10}\left(\frac{E_\textup{max}^{2}}{E_\textup{min}^{2}}\right)
\end{equation}
where $E_\textup{max}$ and $E_\textup{min}$ represent the maximal and minimal amplitudes that the IQ modulator can output in practice. In our work, the extinction ratio of the modulator is high enough to suppress the noise, and $\varepsilon_\textup{Mod}$ can be calculated to be around $1.36\times10^{-4}$ SNU.

The fourth term is phase noise $\varepsilon_\textup{Phase}$, which is also the important part of excess noise. In the LLO CV-QKD scenario,  the phase noise can be divided into two parts, given by
\begin{equation}    \varepsilon_\textup{Phase}=\varepsilon_\textup{fast}+\varepsilon_\textup{slow}
\end{equation}
where $\varepsilon_\textup{fast}$ is the fast-drift phase noise caused by the frequency difference between the Alice's and Bob's lasers, and $\varepsilon_\textup{slow}$ is the slow-drift phase noise caused by the signal transmission in the fiber. In general, $\varepsilon_\textup{fast}=2\pi V_\textup{A}\left(\Delta v_\textup{Alice}+\Delta v_\textup{Bob}\right)/R_\textup{rep}$, herein $R_\textup{rep}$ is the repetition of the signal, $\Delta v_\textup{Alice}$ and  $\Delta v_\textup{Bob}$ are the laser linewidth, respectively. In our work, it can be calculated as around $1\times10^{-4}$ SNU at 80 km and $2.4\times10^{-4}$ SNU at 120 km, respectively. On the other hand, slow drift phase noise is caused by the initial phase difference between the pilot signal and the quantum signal in the channel. When the pilot signal is used to compensate for the quantum signal, an error is introduced, which fluctuates within a certain range. Particularly in long-distance systems, as the distance increases, the compensation accuracy decreases, and the noise increases, thereby affecting the system performance to a greater extent. Moreover, $\varepsilon_\textup{fast}$ and $\varepsilon_\textup{slow}$ are well compensated by the proposed pilot-sequential experiment structure. However, after all phase compensation, there is still residual phase noise. According to $\varepsilon_\textup{rest}=2\eta T V_\textup{A}(1-e^{(-V_\textup{error}/2)})$ \cite{marie2017self}, we can get residual phase noise. First, we calculate the phase of the quantum signal using the X and P orthogonal components after phase compensation. Then, by using the X and P orthogonal components of the quantum signal initially prepared by Alice, we obtain the initial phase information. By calculating the difference between these two phases, we can obtain the residual phase, and from that, determine its variance $V_\textup{error}$. In our experiment, $V_\textup{error}$ for 80 km and 120 km is around 0.005 $\textup{rad}^2$ and $0.35$ $\textup{rad}^2$. The total residual phase noise at 80 km is $0.24\times10^{-3}$ SNU. The total residual phase noise at 120 km is $5.86\times10^{-3}$ SNU. It is clear that the residual phase noise at 120 km is higher. This is primarily due to the accumulation of phase noise and the lower signal-to-noise ratio of the pilot tone at 120 km, which results in a poorer compensation effect.

\begin{figure}[h]
\centering
{\includegraphics[width=\linewidth]{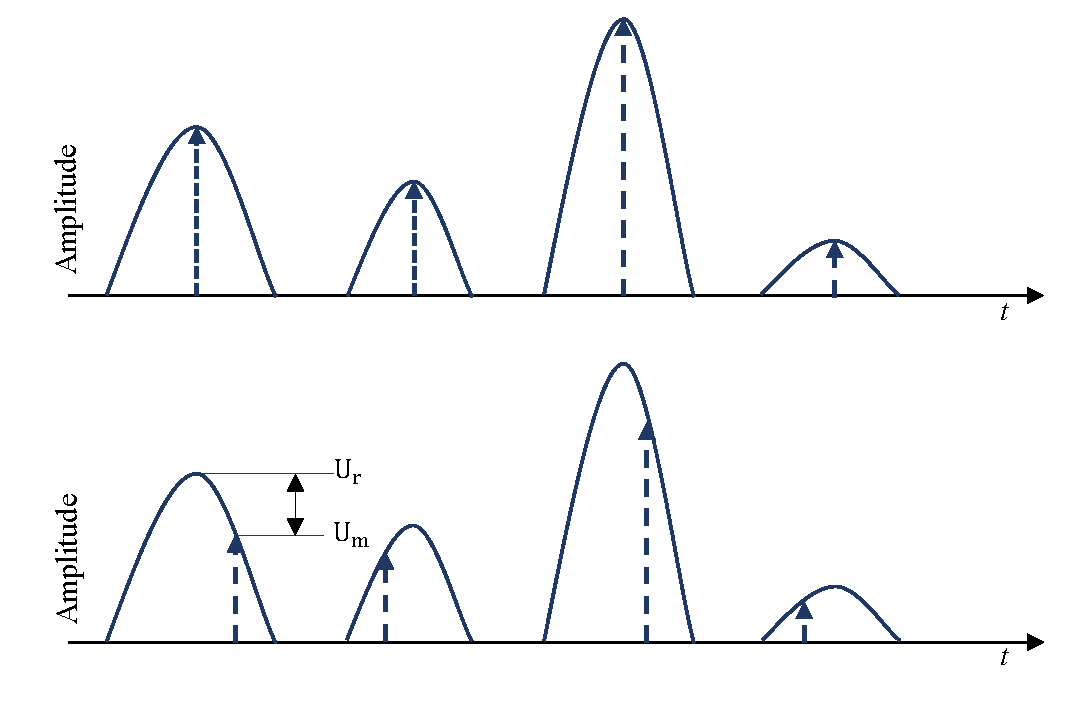}}
\caption{The description of sampling with time jitter between Alice’s and Bob’s clocks (lower), and without time jitter (upper). The arrows denote the sampling times. $\textup{U}_\textup{r}$ and $\textup{U}_\textup{m}$ represent the ideal data and actual sampling data, respectively.}
\label{fig:shizhong}
\end{figure}

In practice, clock synchronization between Alice and Bob is not always stable, and there may be certain drifts in frequency and phase. Crystal oscillators are commonly used as clock sources in digital communication systems, but their frequency may vary due to factors such as temperature changes and power supply fluctuations, leading to frequency jitter in synchronized clocks. In other words, even if the clocks of both parties are perfectly synchronized at the outset, deviations in frequency may occur due to various factors. Over time, these drifts accumulate, resulting in synchronization jitter. However, we often need to sample signals at specific time points, and synchronization jitter can cause instability in sampling times, thereby increasing sampling errors and affecting the accuracy of signal recovery and processing. This phenomenon can be clearly observed from Fig. \ref{fig:shizhong}. Through the DSP mentioned before, it is evident that data sampling occurs at the first of all. Therefore, once clock jitter or frequency drift occurs, it will directly affect subsequent phase compensation, thereby impacting the overall system performance. respectively. The frequency drift caused by the frequency jitter of crystal-controlled oscillators is a random varying value ranging from 0 to 2 ppm. Here in our work, the excess noise caused by imperfect sampling due to this phenomenon can be represented as:
\begin{equation}
    \varepsilon_\textup{sam-error}=Var\left(\textup{U}_\textup{r}-\textup{U}_\textup{m}\right)
    \label{clock 1}
\end{equation}

To better illustrate the impact of this process on the system, a detailed analysis will follow.
First, we define the generated signal at the transmitter as a Gaussian pulse, which can be expressed as:
\begin{equation}
 x(t)=V_p e^{-\frac{t^2}{\delta^2}}
\end{equation}

where $V_p$ is the pulse peak, $\delta^{2}$ denotes the variance of the Gaussian pulse which is $V_A$.    

For a general linear channel, the relationship between the data of Alice and Bob is $y=tx+z$. However, to better focus on analyzing the excess noise introduced by clock jitter, we simplify the relationship between the data of Alice and Bob as $y=tx$, where $t=\sqrt{\eta T}$. By substituting these expressions into Eq.\ref{clock 1}, the excess noise induced by clock jitter can be expanded as
\begin{equation}
\centering
\begin{aligned}
    Var(\textup{U}_\textup{r} - \textup{U}_\textup{m}) &= 
    Var(\sqrt{\eta T_r} x(t) - \sqrt{\eta T_m} x(t + \Delta t)) 
    \\
    &= \eta T_r V_A + \eta T_m V_A^{'} 
    - 2 \eta \sqrt{T_r T_m} \, cov[x(t), x(t + \Delta t)] \\
    &= \eta T_r V_A + \eta T_m V_A^{'} 
    - 2 \rho \eta \sqrt{T_r T_m V_A V_A^{'}} 
\end{aligned}
\label{clock_cal}
\end{equation}

where $T_r$ and $T_m$ denotes the transmittance before and after clock jitter, $\Delta t$ denotes the value of clock jitter, $V_A^{'}$ denotes the variance of the Alice after clock jitter, $\rho$ represents the correlation coefficient of the received data before and after clock jitter and can also equivalently reflect the correlation with the initial Alice's data.

In addition, the relevant theory \cite{chen2023continuous}has already been reported and widely recognized both domestically and internationally, providing theoretical support for this issue. Based the theory of the continuous mode, it can be concluded that sampling errors also lead to a mode mismatch between the transmitter and receiver. By introducing a mode matching coefficient, the received state can be expressed as\cite{chen2023continuous}:
\begin{equation}
    \hat{A}_{\Xi}^{\dagger}=\sqrt{\eta_{m}} \hat{A}_{\xi}^{\dagger}+\sqrt{1-\eta_{m}} \hat{A}_{\Psi_{\perp}}^{\dagger}
\end{equation}
where $\eta_{m}=\left[\int d t \Xi^{*}(t) \xi(t)\right]^{2} \leq 1$ represents the mode-matching coefficient.
Based on the channel noise introduced by channel loss \cite{RevModPhys.81.1301}, the noise caused by mode mismatch is defined as:
\begin{equation}
    \varepsilon_{m}=\frac{1-\eta_{m}}{\eta_{m}}
    \label{match}
\end{equation}

Eq. \ref{match} provides a theoretical foundation for the current experimental issue and deep research on this problem using continuous mode is already underway.

These are basically the components of excess noise. Combing the proposed technologies, we can suppress excess noise to a lower level in the experiment which can be seen in the following section. 

\subsection{Experimental Results}
\begin{figure}[!t]
		\begin{overpic}[width=0.45\textwidth]{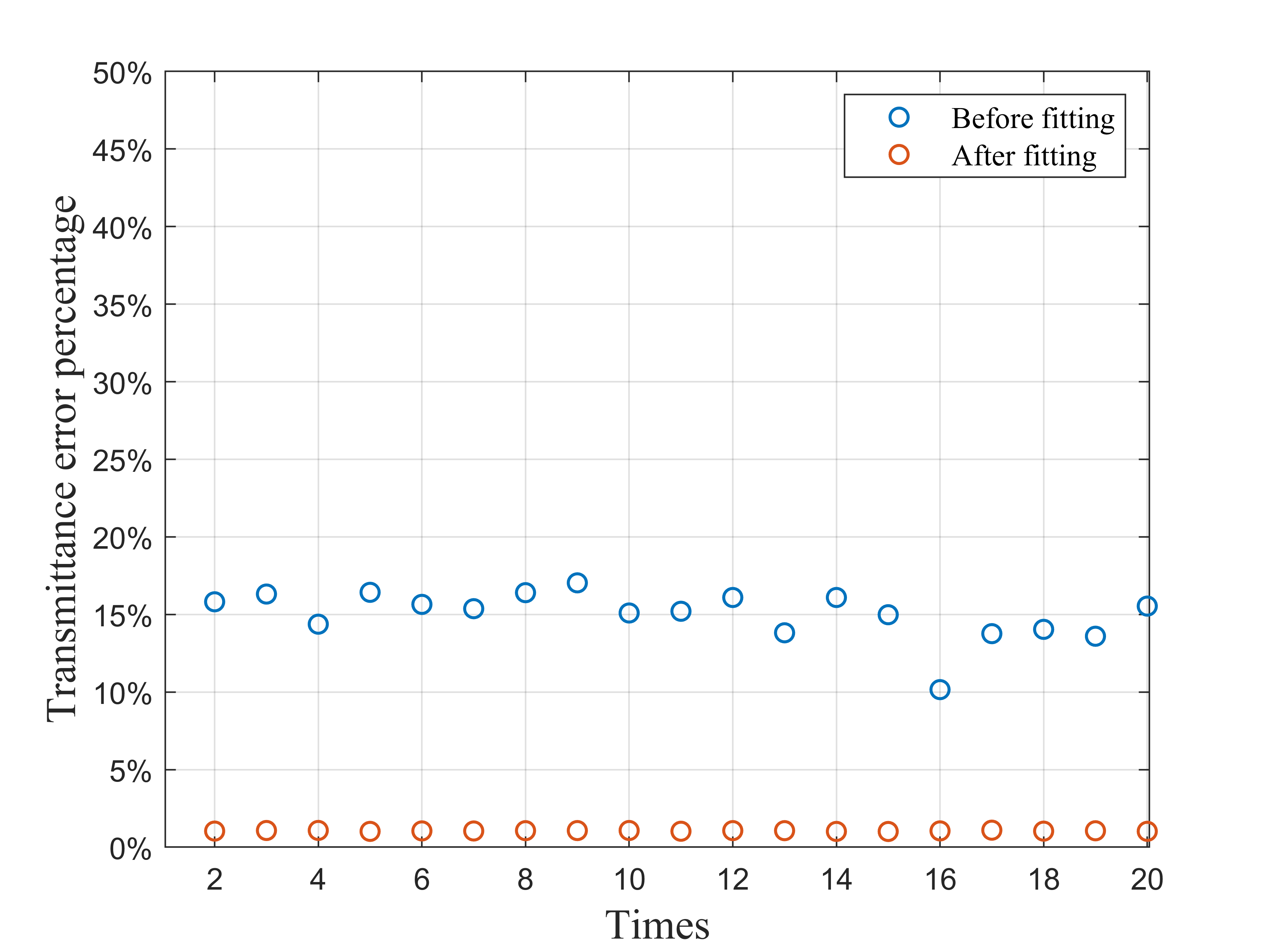}
			\put(-1,70){\large\textbf{(a)}}
		\end{overpic}
		\hspace{2mm}
        
		\begin{overpic}[width=0.45\textwidth]{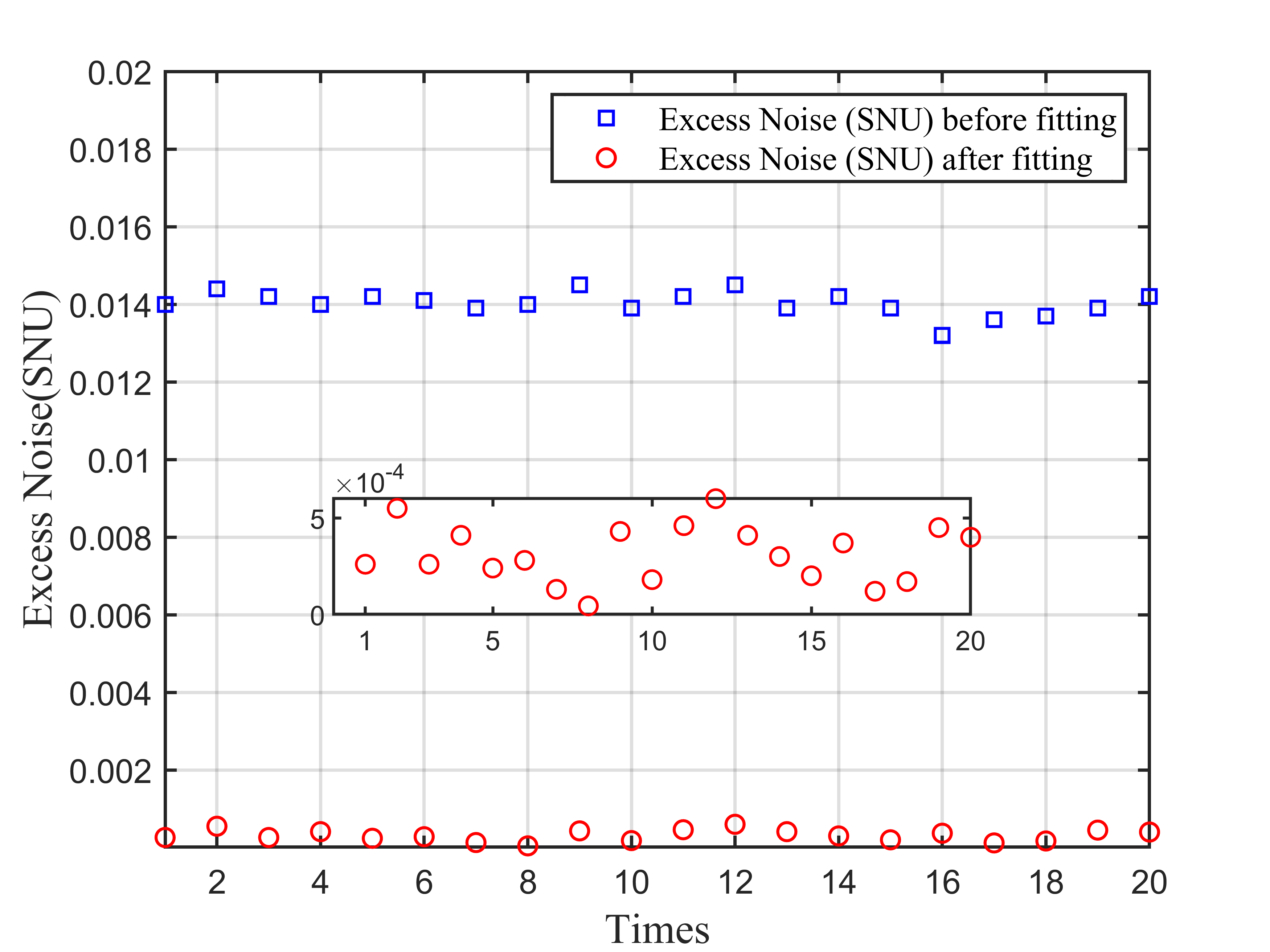}
			\put(-1,70){\large\textbf{(b)}}
		\end{overpic}
		\hspace{2mm}
		\begin{overpic}[width=0.45\textwidth]{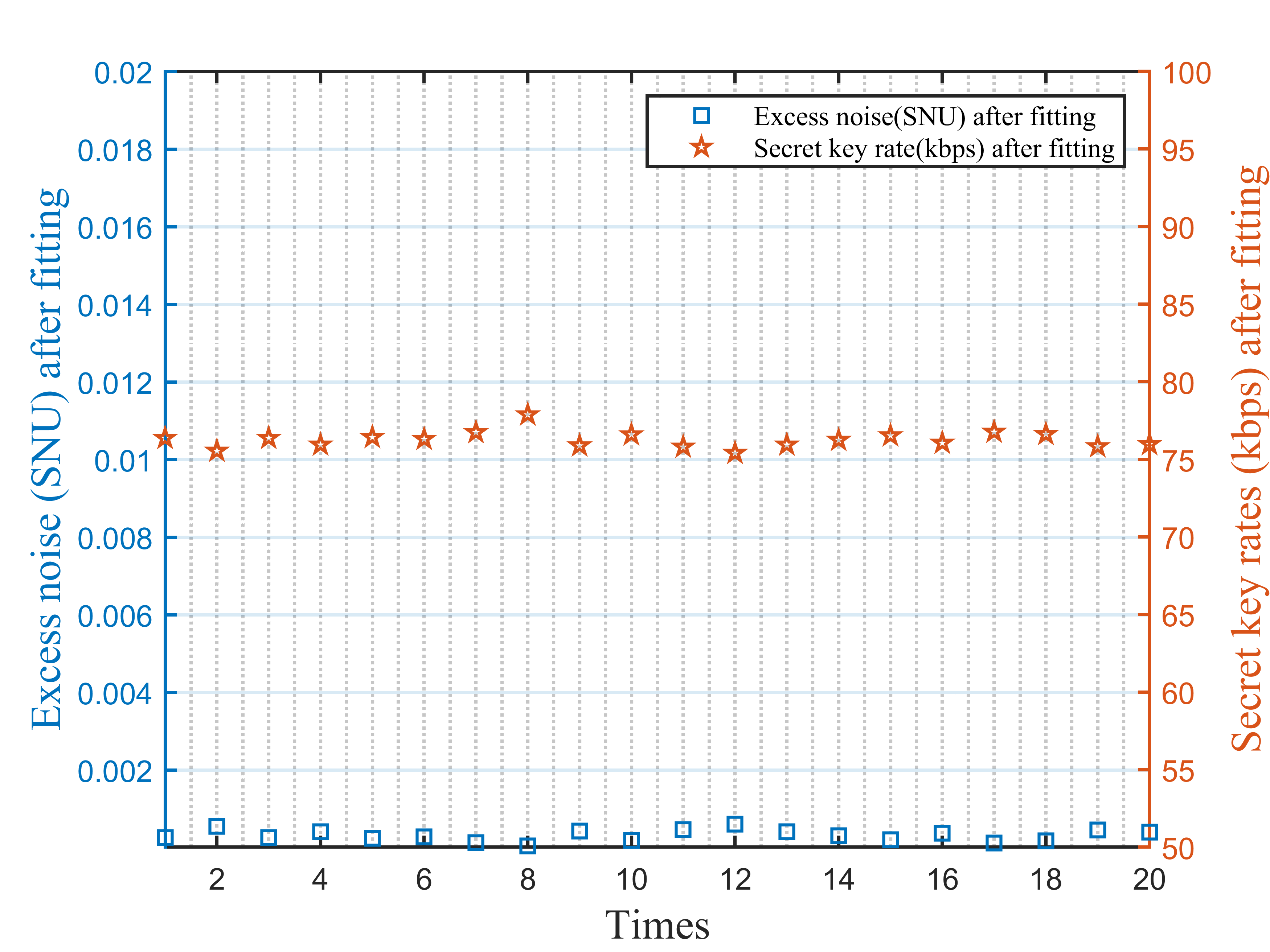}
			\put(-1,70){\large\textbf{(c)}}
		\end{overpic}
		\captionsetup[subfigure]{labelformat=empty}  
\caption{Comparison of experimental results before and after least squares fitting over 80 km. 
(\textbf{a}) Changes in transmittance error percentage before and after LS algorithm at 80 km.
(\textbf{b}) Experimental excess noise (SNU) levels before and after least squares fitting.  
(\textbf{c}) Experimental secret key rate and excess noise (SNU) levels after least squares fitting.}
\label{fig:80km}
\end{figure}

The experimental results of the LLO CV-QKD over80 km and 120 km are presented in the following. Firstly, the performance of LLO CV-QKD setup over 80 km was shown as follows. Fig. \ref{fig:80km} shows the experimental excess noise over 1 hour and the corresponding secret key rate. It is worth noting that we significantly reduced the data acquisition time by using continuous acquisition card. Based on Fig. \ref{fig:shizhong} and the corresponding theoretical analysis, it is clear that sampling error leads to transmittance errors. To better analyze the effectiveness of LS in compensating for these errors, Fig. \ref{fig:80km}(a) presents an analysis of the transmittance error percentage before and after applying LS.The blue and red circles represent the results before and after using LS. From Fig. \ref{fig:80km}(a), we can clearly observe that before using LS, the estimation error of transmittance is around 15\%, whereas after using LS, the transmittance error is consistently controlled at a level of 1\% or even lower. The comparison before and after clearly shows that LS has a good effect on calibrating sampling errors. In Fig. \ref{fig:80km} (b), the blue square markers above represents the excess noise before fitting, while the red circular markers below represents the excess noise after fitting. It is evident that the least squares fitting algorithm greatly reduces the additional excess noise introduced by data bias mentioned earlier. Since the level of excess noise after fitting is low, we have included an enlarged image in the middle for more intuitive observation. The excess noise after fitting is at the level of milli shot noise (mSNU) which is already quite low. The mean excess noise before and after fitting are 0.0139 SNU and 0.38 mSNU. These results demonstrate the effectiveness of our proposed algorithm and the stability of the experimental system, as it can achieve long-term stability with low excess noise.

\graphicspath{{Figures/}{logo/}}
 
To taking finite-size effects into account, the secret key rate is given by the general formula \cite{leverrier2010finite}:
\begin{equation}
 K=f_\textup{sym}\cdot\frac{n}{N}(1-FER)(\beta I_\textup{AB}-\chi_\textup{BE}-\Delta(n))
\end{equation}
where $f_\textup{sym}$ is the repetition rate of the quantum signal, $N$ represents the total number of a block while $n$ represents the number for key calculation, $FER$ is the frame error rate of the reconciliation, $\beta$ is the reconciliation efficiency in the post-processing stage, $I_\textup{AB}$ is the mutual information between legitimate users Alice and Bob, $\chi_\textup{BE}$ is the Holevo bound and $\Delta(n)$ is the secret key rate decrease caused by finite-size block effect.

\begin{figure}[h]
\centering
{\includegraphics[width=\linewidth]{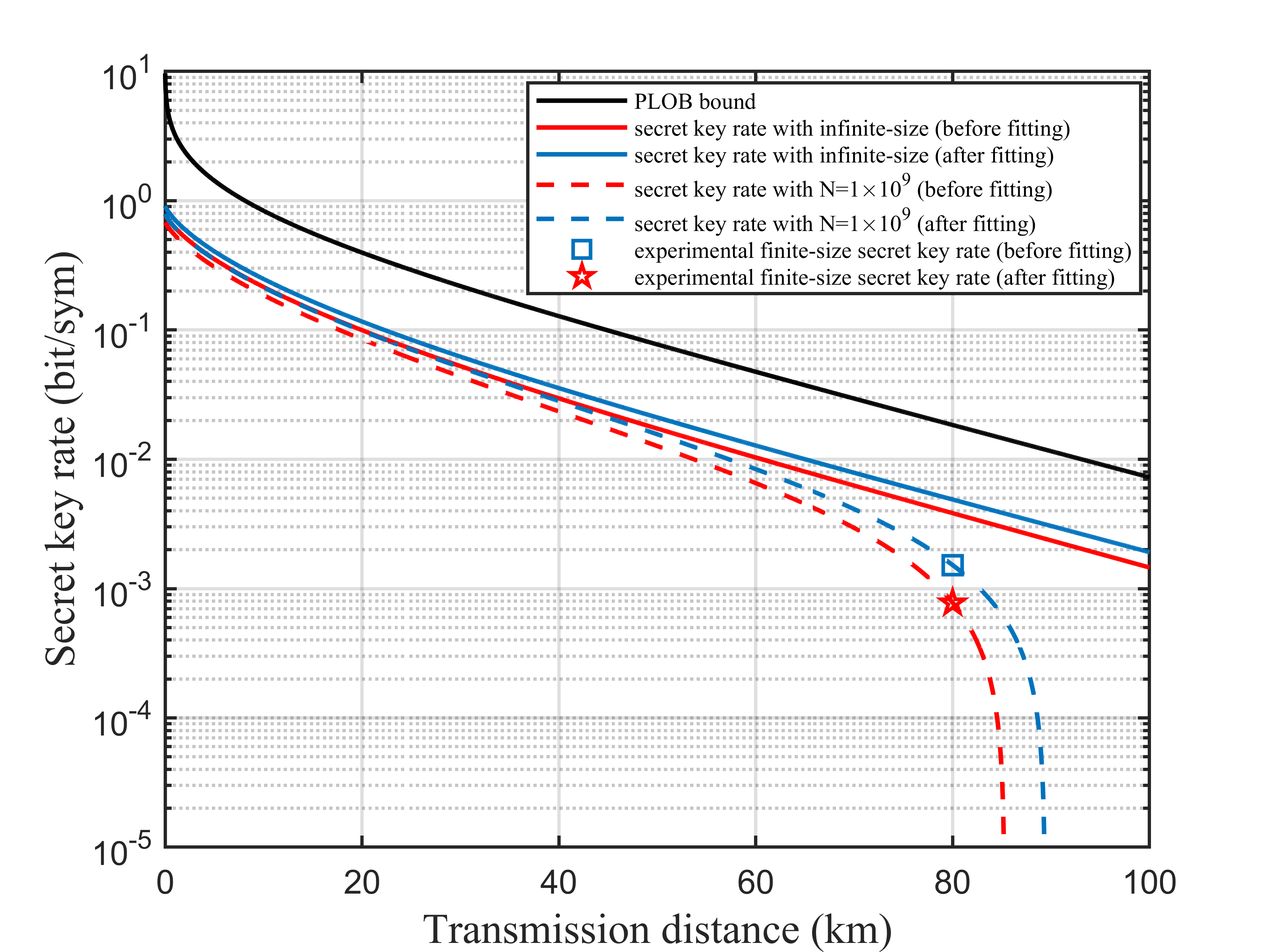}}
\caption{Secret key rates curves of experiment as a function of transmission distance.
The black solid line represents the PLOB bound in this scheme. The solid line in different colors represent the secret key rate before and after least squares fitting in infinite-size scenarios, respectively. The dash line in different colors represent the secret key rate before and after least squares fitting under under finite-size block of $1\times10^{9}$, respectively. The diamond and the pentagram represents experimental secret key rate before and after least squares fitting with block size of $1\times10^{9}$, respectively. 
}
\label{fig:80kmskr}
\end{figure}

Immediately after, we provide the experimental excess noise and the corresponding secret key rate with block size of $1\times10^{9}$ in Fig. \ref{fig:80km} (c). The pentagon markers above represent the experimental secret key rate, while the square markers below represent the experimental excess noise. It can be seen that we can get a stable and relatively higher secret key rate. Furthermore, the secret key rates of both numerical simulations and experimental results are depicted in Fig. \ref{fig:80kmskr}. The black solid line represents the PLOB \cite{pirandola2017fundamental} bound of GMCS CV-QKD protocol with heterodyne detection. The blue and red solid line represent the secret key rate before and after least squares fitting in infinite-size scenarios, respectively. In contrast, the blue and red dash line is the secret key rate before and after least squares fitting with block size of $1\times10^9$. The square and pentagon correspond to our experimental results before and after least squares fitting taking finite-size effects into account, respectively. The secret key rate before fitting is 38.93 kbps, while the secret key rate after fitting is 76 kbps. The results clearly show that we have significantly increased the secret key rate. Furthermore, we can also observe that a longer secure distance can be got under the same block size. This also presents a new approach for long-distance transmission in the future. Table \ref{tab:shape-functions} outlines the parameters employed in the calculation of the secret key.
\begin{table*}[htbp]
\begin{center}
\caption{{\bf Overview of experimental parameters for different fiber lengths.} L, fiber length; $\beta$, reconciliation efficiency, $FER$, frame error rate of the reconciliation, $V_\textup{A}$, modulation variance; $\eta$, detection efficiency; $v_\textup{el}$, electric noise; $\varepsilon_\textup{1}$, excess noise before fitting; $\varepsilon_\textup{2}$, excess noise after fitting.}
  \begin{tabular}{c c c c c c c c}
  \toprule[1.2pt]
  \toprule[1.2pt]
    \makebox[0.1\textwidth][c]{L} & \makebox[0.1\textwidth][c]{$\beta$} & \makebox[0.1\textwidth][c]{$FER$}
    & \makebox[0.1\textwidth][c]{$V_\textup{A}$}
    & \makebox[0.1\textwidth][c]{$\eta$} & \makebox[0.1\textwidth][c]{$v_\textup{el}$}& \makebox[0.1\textwidth][c]{$\varepsilon_\textup{1}$}& \makebox[0.1\textwidth][c]{$\varepsilon_\textup{2}$}       \\
    \midrule[1pt]
80 km & 95\% & 0.1 & 4.01 SNU & 0.481 & 0.0372 SNU & 0.014 SNU & 0.38 mSNU\\
120 km & 95\% & 0.1& 9.41 SNU & 0.481 & 0.0374 SNU & 0.022 SNU & 0.006 SNU\\
\bottomrule[1pt]
\end{tabular}
\label{tab:shape-functions}
\end{center}
\end{table*}

\begin{figure}[!t]
		\begin{overpic}[width=0.45\textwidth]{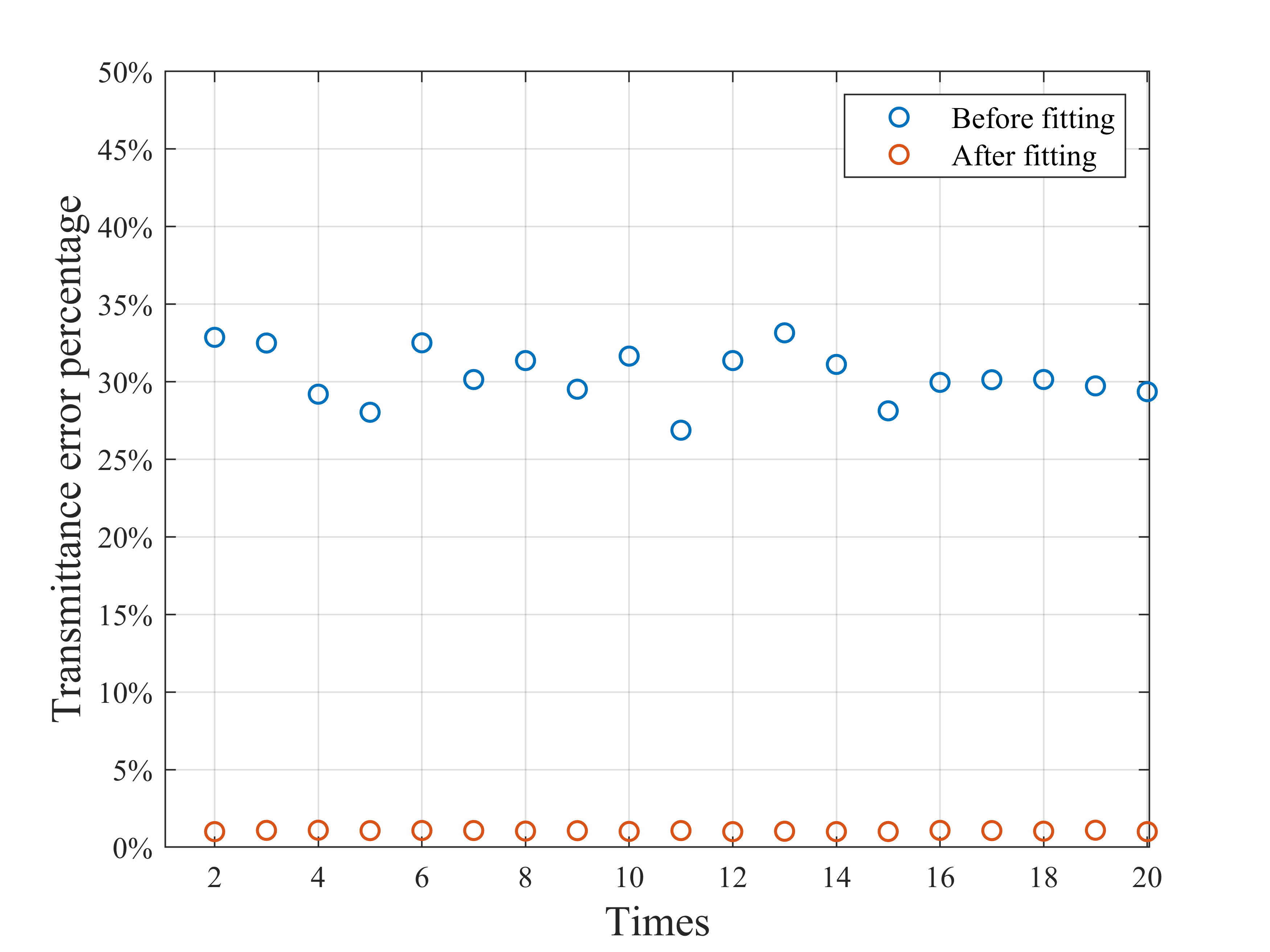}
		\put(-1,71){\large\textbf{(a)}}
		\end{overpic}
		\hspace{2mm}
        
		\begin{overpic}[width=0.45\textwidth]{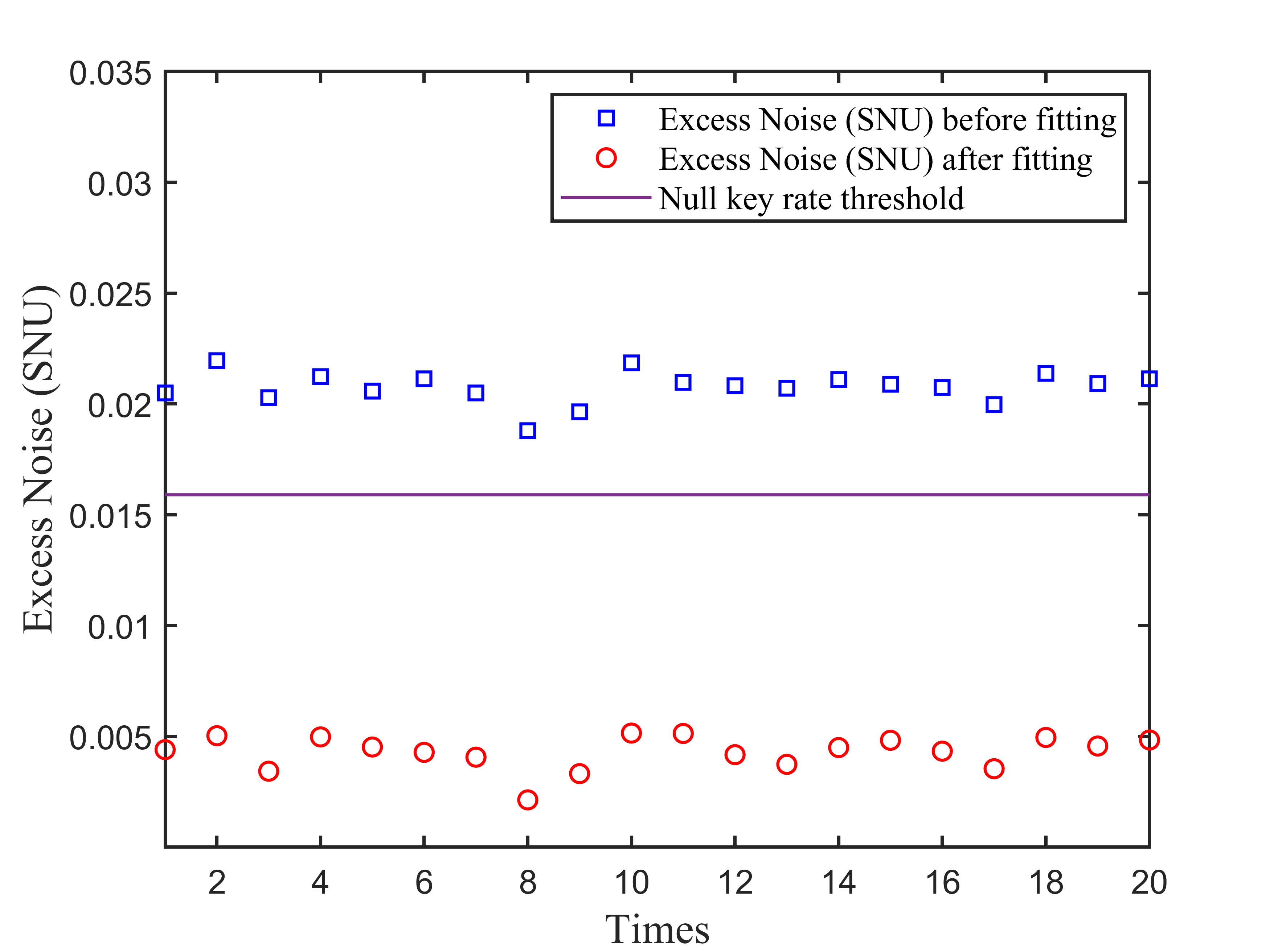}
		\put(-1,71){\large\textbf{(b)}}
		\end{overpic}
		\hspace{2mm}
		\begin{overpic}[width=0.45\textwidth]{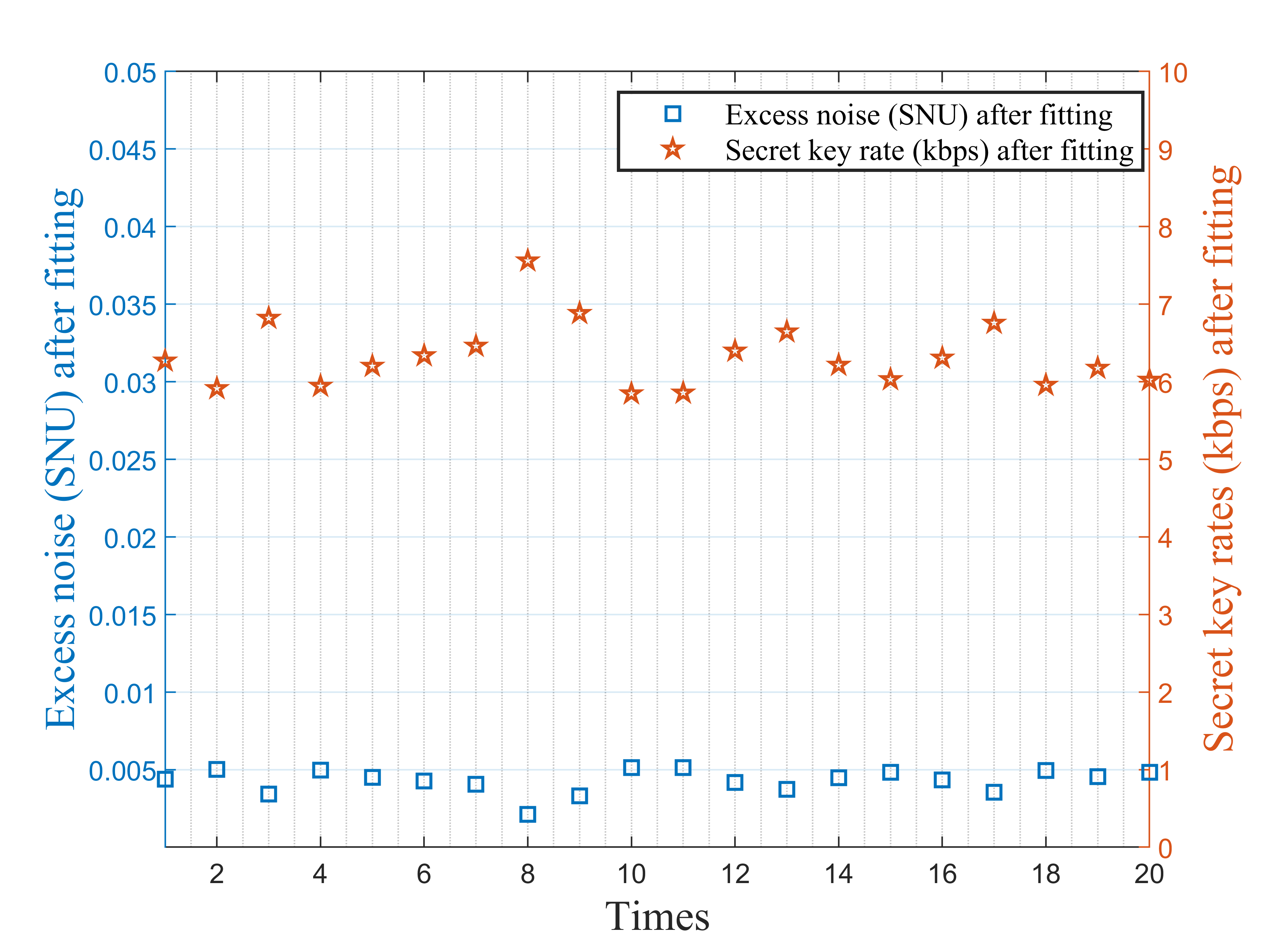}
		\put(-1,71){\large\textbf{(c)}}
		\end{overpic}
		\captionsetup[subfigure]{labelformat=empty}  
\caption{Comparison of experimental results before and after least squares fitting over 120 km.
(\textbf{a}) Changes in transmittance error percentage before and after LS algorithm at 120 km.
(\textbf{b}) Experimental excess noise (SNU) levels before and after least squares fitting. The purple line indicates the maximal value of excess noise that allows for a positive secret key rate.
(\textbf{c}) Experimental secret key rate and excess noise (SNU) levels after least squares fitting.}
\label{fig:120km}
\end{figure}

Secondly, we will show the performance of LLO CV-QKD setup over 120 km as follows. It is the longest experimental validation to date of LLO CV-QKD, ensuring secret key generation while considering finite-size effects. Fig. \ref{fig:120km} shows the experimental excess noise and the corresponding secret key rate  for a distance of 120 km. In Fig. \ref{fig:120km}(a) presents an analysis of the transmittance error percentage before and after applying LS. The blue and red circles represent the results before and after using LS. From Fig. \ref{fig:120km}(a), we can clearly observe that before using LS, the estimation error of transmittance is around 30\%, whereas after using LS, the transmittance error is consistently controlled at a level of 1\% or even lower. Note that, the transmittance error at 120 km is larger than that at 80 km, which is due to the fact that clock jitter is a random and cumulative process. Longer distances require larger finite-size block, which means more accumulated bias. In Fig. \ref{fig:120km} (b), the blue square markers above represents the excess noise before fitting, while the red circular markers below represents the excess noise after fitting. In addition, the purple line indicates the maximal value of excess noise that allows for a positive secret key rate. It's evident that without enhancing the data recovery of the received signal, no secret key can be extracted. The mean excess noise before and after fitting are 0.022 SNU and 0.006 SNU. Next, we provide the experimental excess noise and the corresponding secret key rate with block size of $1\times10^{11}$ in Fig. \ref{fig:120km} (c).

\graphicspath{{Figures/}{logo/}}

\begin{figure}[h]
\centering
{\includegraphics[width=\linewidth]{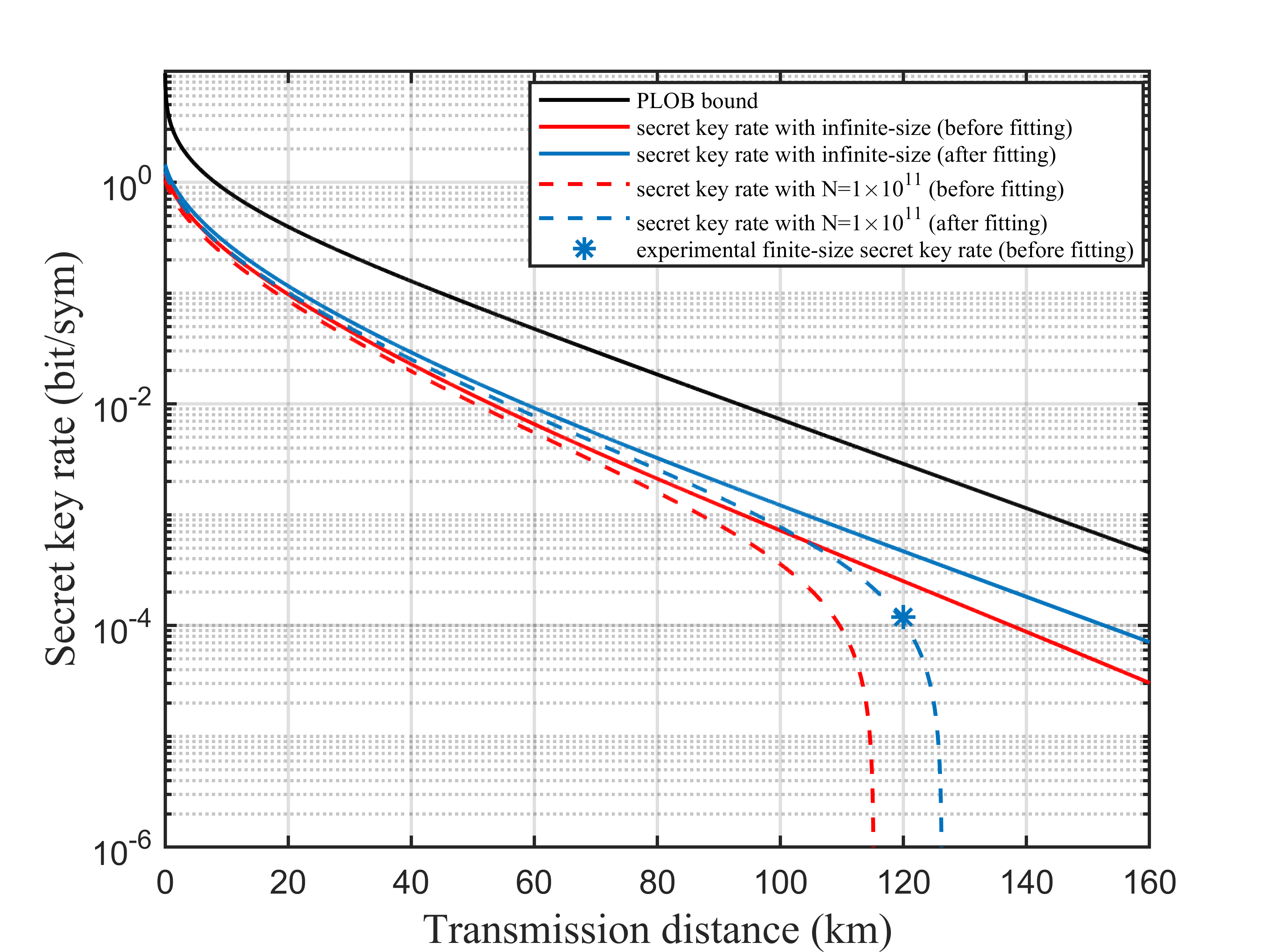}}
\caption{Secret key rates curves of experiment as a function of transmission distance.
The black solid line represents the PLOB bound in this scheme. The solid line in different colors represent the secret key rate before and after least squares fitting in infinite-size scenarios, respectively. The dash line in different colors represent the secret key rate before and after least squares fitting under under finite-size block of $1\times10^{11}$, respectively. The diamond and the pentagram represents experimental secret key rate before and after least squares fitting with block size of $1\times10^{11}$, respectively. 
}
\label{fig:120kmskr}
\end{figure}

The secret key rates of both numerical simulations and experimental results are depicted in Fig. \ref{fig:120kmskr}. The blue and red solid line represent the secret key rate before and after least squares fitting in infinite-size scenarios, respectively. In contrast, the blue and red dash line is the secret key rate before and after least squares fitting with block size of $1\times10^{11}$. The square and pentagon correspond to our experimental results before and after least squares fitting taking finite-size effects into account, respectively. It's evident that we break further distances than before after the least squares fitting from the results. The final secret key rate is 5.36 kbps at 120 km. The experimental key parameters also can be found in Table \ref{tab:shape-functions}.

\section{Discussion}

\begin{figure}[h]
\centering
{\includegraphics[width=\linewidth]{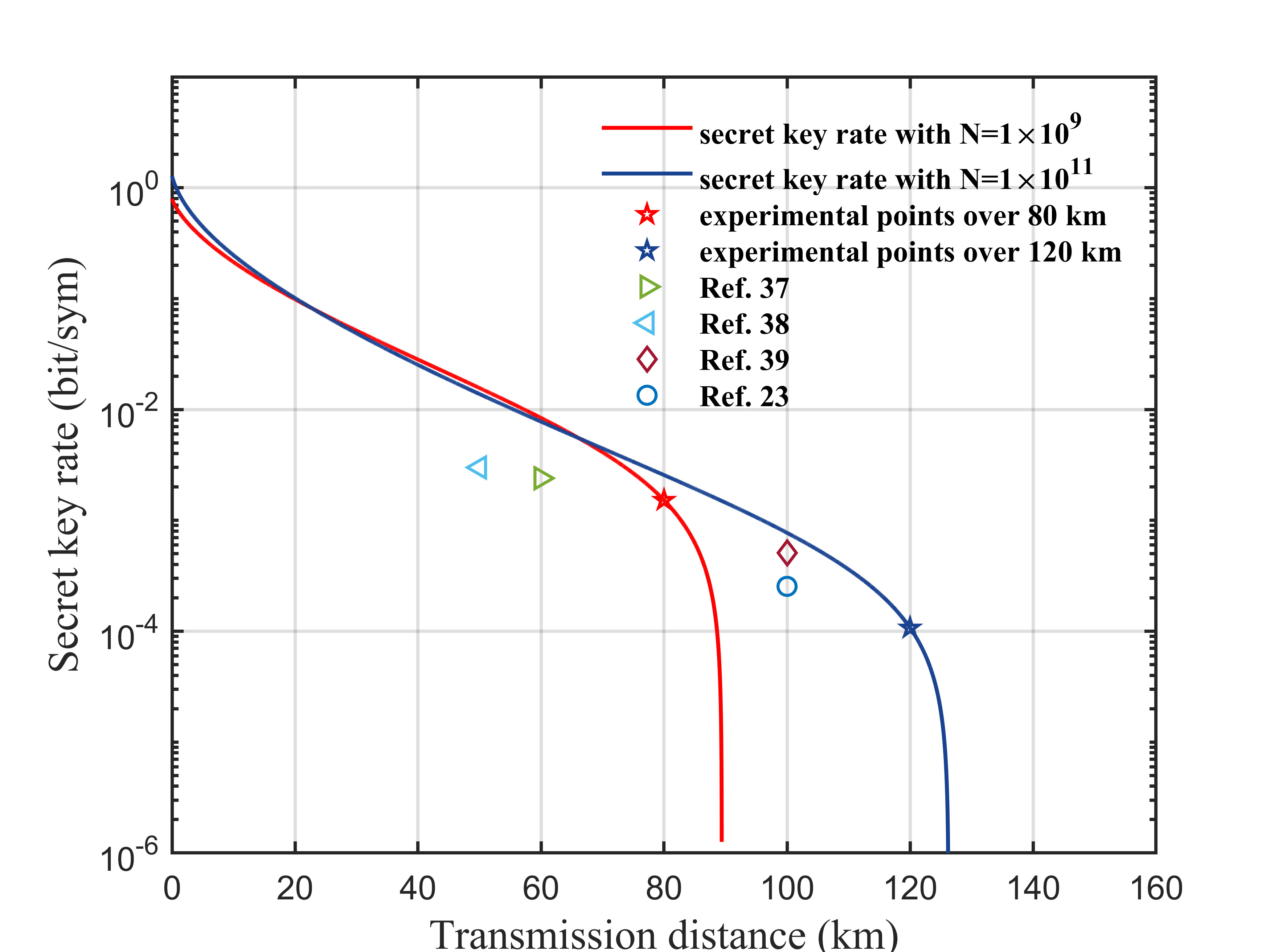}}
\caption{Experimental key rates in our work and previous state-of-the-art experimental results \cite{hajomer2022continuous,li2023continuous,pi2023sub,hajomer2024long} for comparison.
}
\label{fig:skrduibi}
\end{figure}

In recent years, several progress has been made in long-distance CV-QKD LLO experiments. For example, the first CV-QKD experiment that enables the generation of secure key over a 60 km fiber channel with locally generated local oscillator was reported in \cite{hajomer2022continuous}. Moreover, \cite{li2023continuous} designed and fabricated two on-chip tunable lasers for CV-QKD, and demonstrated a high-performance system based on these sources over 50 km fiber distance, which marks a breakthrough toward building a fully integrated CV-QKD. Furthermore, a sub-Mbps key rate LLO CV-QKD over a 100-km transmission distance was successfully reported in \cite{pi2023sub}by designing a high accuracy data-assisted time domain equalization algorithm to compensate the phase noise and polarization variation in low signal-to-noise ratio. However, they do not consider the effect of finite-size block, which is very important in practice and cannot be ignored. Recently, the experiment \cite{hajomer2024long} has claimed they have successfully demonstrated LLO CV-QKD over 100 km of ultralow-loss optical fiber by controlling the phase noise–induced excess noise through a machine learning framework for carrier recovery and optimizing the modulation variance. In our work, we have successfully demonstrated a GMCS LLO CV-QKD experiment over 80 km and 120 km which is the longest distance to date. For both distances, we use standard optical fiber and take into account the effect of the actual finite-size block. For comparison, the previously reported experimental results are also shown in Fig. \ref{fig:skrduibi}. 

Taking into account factors such as the influence of clock jitter on clock synchronization between Alice and Bob, leading to deviations in data sampling and the difficulties in data recovery at low signal-to-noise ratios, all these factors ultimately contribute to an increase in overall excess noise in the system. Therefore, we defined this additional introduced noise as sampling-error noise and completed its quantification and analysis. Considering the lower tolerance for excess noise over long distances, we proposed a new approach: the least squares fitting algorithm to preprocess the collected signals, obtaining data closer to the true values and facilitating subsequent signal processing. Thus, the excess noise at mSNU level could be obtained. Additionally, we designed high efficient post-processing algorithms to significantly improve real-time processing speed and more efficiently generate secret key rate in real-time. Finally,we realized a secret key rate of 76 kbps under transmission distance of 80 km with block size of $1\times10^{9}$ and also achieved a secret key rate of 5.36 kbps under transmission distance of 120 km with block size of $1\times10^{11}$. Our work provides a feasible way of recovering accurate data under low signal-to-noise ratio. It can also break further distances and pave the way for large-scale quantum secure communications in the future. While as the transmission distance increases, the lower signal-to-noise ratio will make data recovery more complex. In the future work, we will further optimize algorithms to adapt to different transmission conditions and improve the robustness of the system.

\appendix
\renewcommand{\thesection}{APPENDIX \Alph{section}}

\section{Digital signal processing algorithm}
\label{app:a}
First, The quantum signal with GMCS and the pilot tone can be given by
\begin{equation}
 \renewcommand{\theequation}{A\arabic{equation}}
  \setcounter{equation}{1}
E_\textup{sig}=A_\textup{sig}\textup{cos}\left(2\pi f_\textup{A}t+\varphi_\textup{A}+\varphi_\textup{sig}\right)
\end{equation}

\begin{equation}
 \renewcommand{\theequation}{A\arabic{equation}}
  \setcounter{equation}{2}
E_\textup{pilot}=A_\textup{pilot}\textup{cos}\left(2\pi f_\textup{A}t+\varphi_\textup{A}\right)
\end{equation}
where $A_\textup{sig}$ is the amplitude of the quantum signals and $\varphi_\textup{sig}$ is the initial phase of the quantum signals f0llow the Rayleigh distribution and the uniform distribution respectively, $f_\textup{A}$ denotes the central frequency of the Alice's laser, $\varphi_\textup{A}$ denotes the initial phase of Alice's laser, $A_\textup{pilot}$ represents the amplitude of the pilot tone.
At Bob's side, the LO can be expressed by
\begin{equation}
 \renewcommand{\theequation}{A\arabic{equation}}
  \setcounter{equation}{3}
E_\textup{LO}=A_\textup{LO}\textup{cos}\left(2\pi f_\textup{B}t+\varphi_\textup{B}\right)
\end{equation}
where $A_\textup{LO}$ denotes the amplitude of the LO, $f_\textup{B}$ denotes the central frequency of the Bob's laser and $\varphi_\textup{B}$ denotes the initial phase of Bob's laser.

After heterodyne detection, the generated photocurrents can be expressed as, respectively

\begin{equation}
 \renewcommand{\theequation}{A\arabic{equation}}
  \setcounter{equation}{4}
\begin{aligned}
I_\textup{sig}&=\eta\left(\lvert E_\textup{sig}+E_\textup{LO} \rvert^2-\lvert E_\textup{sig}-E_\textup{LO} \rvert^2\right)
\\
&=2\eta \text{A}_\textup{sig}\text{A}_\textup{LO}\textup{cos}\left(2\pi\Delta f_\textup{AB}t+\varphi_\textup{sig}+\Delta\varphi_\textup{fast}+\Delta\varphi_\textup{slow}\right )    
\end{aligned}
\end{equation}

\begin{equation}
 \renewcommand{\theequation}{A\arabic{equation}}
  \setcounter{equation}{5}
\begin{aligned}
I_\textup{pilot}&=\eta\left(\lvert E_\textup{pilot}+E_\textup{LO} \rvert^2-\lvert E_\textup{pilot}-E_\textup{LO} \rvert^2\right)
\\
&=2\eta \text{A}_\textup{pilot}\text{A}_\textup{LO}\textup{cos}\left(2\pi\Delta f_\textup{AB}t+\Delta\varphi_\textup{fast}\right )
\end{aligned}
\end{equation}
with the frequency difference of two lasers $\Delta f_\textup{AB}=f_\textup{A}-f_\textup{B}$, the fast-drift laser phase $\Delta\varphi_\textup{fast}=\varphi_\textup{A}-\varphi_\textup{B}$ and the slow-drift channel phase $\Delta\varphi_\textup{slow}=\varphi_\textup{sig}^\textup{channel}-\varphi_\textup{pilot}^\textup{channel}$, in which $\varphi_\textup{sig}^\textup{channel}$ and $\varphi_\textup{{pilot}}^\textup{channel}$ denotes the the phase drift of the quantum signal and pilot tone distributed by the quantum channel, respectively, where $\eta$ denote the detection efficiency of the BHD. 

Next, we perform orthogonal down-conversion and low-pass filtering to obtain the orthogonal components of both the signal and the pilot, as follows
\begin{equation}
 \renewcommand{\theequation}{A\arabic{equation}}
  \setcounter{equation}{6}
\begin{aligned}
\textup{X}_\textup{sig}&=LPF\left(real\left(I_\textup{sig}\cdot e^{-2\pi\Delta f_\textup{AB} t}\right)\right)
\\
&=\eta \text{A}_\textup{sig}\text{A}_\textup{LO}\textup{cos}\left(\varphi_\textup{sig}+\Delta\varphi_\textup{fast}+\Delta\varphi_\textup{slow}\right)    
\end{aligned} 
\end{equation}

\begin{equation}
 \renewcommand{\theequation}{A\arabic{equation}}
  \setcounter{equation}{7}
\begin{aligned}
\textup{P}_\textup{sig}&=LPF\left(imag\left(I_\textup{sig}\cdot e^{-2\pi\Delta f_\textup{AB} t}\right)\right)
\\
&=-\eta \text{A}_\textup{sig}\text{A}_\textup{LO}\textup{sin}\left(\varphi_\textup{sig}+\Delta\varphi_\textup{fast}+\Delta\varphi_\textup{slow}\right)    
\end{aligned} 
\end{equation}

\begin{equation}
 \renewcommand{\theequation}{A\arabic{equation}}
  \setcounter{equation}{8}
\begin{aligned}
\textup{X}_\textup{pilot}&=LPF\left(real\left(I_\textup{pilot}\cdot e^{-2\pi\Delta f_\textup{AB} t}\right)\right)
\\
&=\eta \text{A}_\textup{pilot}\text{A}_\textup{LO}\textup{cos}\left(\Delta\varphi_\textup{fast}\right)  
\end{aligned} 
\end{equation}

\begin{equation}
 \renewcommand{\theequation}{A\arabic{equation}}
  \setcounter{equation}{9}
\begin{aligned}
\textup{P}_\textup{pilot}&=LPF\left(imag\left(I_\textup{pilot}\cdot e^{-2\pi\Delta f_\textup{AB} t}\right)\right)
\\
&=-\eta \text{A}_\textup{pilot}\text{A}_\textup{LO}\textup{sin}\left(\Delta\varphi_\textup{fast}\right)  
\end{aligned} 
\end{equation}
where $LPF\left(\cdot\right)$ represents the function of low-pass filtering, $real\left(\cdot\right)$ denotes the function to obtain the real part X of orthogonal components while $imag\left(\cdot\right)$ denotes the function to obtain the imaginary part P of orthogonal components. In our experiment, we use a Butterworth low-pass filter with a bandwidth of approximately 100 MHz, to filter out both the quantum signal and the pilot tone signal, with an SNR $\approx -3.74 \textup{dB}$ at 80 km and $\approx -8.08 \textup{dB}$ at 120 km,respectively.

Next, we utilized the X and P components of the pilot tone to obtain the phase drift angle corresponding to the fast drift, can be given by
\begin{equation}
 \renewcommand{\theequation}{A\arabic{equation}}
  \setcounter{equation}{10}
\begin{aligned}
\theta_\textup{fast}=-\textup{arctan}\frac{\textup{P}_\textup{pilot}}{\textup{X}_\textup{pilot}}  
\end{aligned}
\end{equation}
Afterwards, we utilized the calculated angle to execute the corresponding fast phase drift compensation as follows 
\begin{equation}
 \renewcommand{\theequation}{A\arabic{equation}}
  \setcounter{equation}{11}
\begin{aligned}
\textup{X}_\textup{sig}^{'}&=\textup{X}_\textup{sig}\textup{cos}\theta_\textup{fast}-\textup{P}_\textup{sig}\textup{sin}\theta_\textup{fast}
\\
&=\eta
\text{A}_\textup{sig}\text{A}_\textup{LO}\textup{cos}\left(\varphi_\textup{sig}+\Delta\varphi_\textup{slow}\right)   
\end{aligned} 
\end{equation}

\begin{equation}
 \renewcommand{\theequation}{A\arabic{equation}}
  \setcounter{equation}{12}
\begin{aligned}
\textup{P}_\textup{sig}^{'}&=-\textup{X}_\textup{sig}\textup{sin}\theta_\textup{fast}-\textup{P}_\textup{sig}\textup{cos}\theta_\textup{fast}
\\
&=-\eta
\text{A}_\textup{sig}\text{A}_\textup{LO}\textup{sin}\left(\varphi_\textup{sig}+\Delta\varphi_\textup{slow}\right)   
\end{aligned} 
\end{equation}

After LS, because the phase of the noise is randomly distributed, we use a phase searching algorithm: iterating over angles ranging from 0 to $2\pi$, computing the correlation between the compensated results and the initial Alice's data. The angle $\theta_\textup{slow}$ corresponding to the maximum correlation is then chosen. The received quadrature components $X_\textup{B}$ and $P_\textup{B}$ can be more precisely compensated as
\begin{equation}
 \renewcommand{\theequation}{A\arabic{equation}}
  \setcounter{equation}{13}
\begin{aligned}
\textup{X}_\textup{B}&=\textup{X}_\textup{sig}^{'}\textup{cos}\theta_\textup{slow}-\textup{P}_\textup{sig}^{'}\textup{sin}\theta_\textup{slow}
\\
&=\eta
\text{A}_\textup{sig}\text{A}_\textup{LO}\textup{cos}\left(\varphi_\textup{sig}\right)   
\end{aligned} 
\end{equation}

\begin{equation}
 \renewcommand{\theequation}{A\arabic{equation}}
  \setcounter{equation}{14}
\begin{aligned}
\textup{P}_\textup{B}&=-\textup{X}_\textup{sig}^{'}\textup{sin}\theta_\textup{slow}-\textup{P}_\textup{sig}^{'}\textup{cos}\theta_\textup{slow}
\\
&=-\eta
\text{A}_\textup{sig}\text{A}_\textup{LO}\textup{sin}\left(\varphi_\textup{sig}\right)   
\end{aligned} 
\end{equation}

It is worth noting that since those data used for estimating $\theta_\textup{slow}$ will eventually be discarded, the security of this process is also guaranteed\cite{jouguet2013experimental,soh2015self,qi2015generating,wang2018high3,chin2021machine,hajomer2024long}.

\section{Secret Key Rate Calculation}
\label{app:b}
The secret key rate considering the finite-size block effect can be written as\cite{leverrier2010finite}
\begin{equation}
 \renewcommand{\theequation}{B\arabic{equation}}
 \setcounter{equation}{1}
K=f_\textup{sym}\cdot\frac{n}{N}(1-FER)(\beta I_\textup{AB}-\chi_\textup{BE}-\Delta(n)),
\end{equation}
where $f_\textup{sym}$ is the repetition rate of the quantum signal, $N$ represents the total number of a block while $n$ represents the number for key calculation, $FER$ is the frame error rate of the reconciliation, $\beta$ is the reconciliation efficiency in the post-processing stage, $I_\textup{AB}$ is the mutual information between legitimate users Alice and Bob, and~$\chi_\textup{BE}$ is the Holevo bound. $I_\textup{AB}$ is represented by
\begin{equation}
 \renewcommand{\theequation}{B\arabic{equation}}
 \setcounter{equation}{2}
I_\textup{AB}= \text {log }_{2}\left(\frac{V+\chi_{\text{tot}}}{1+\chi_{\text{tot}}}\right),
\end{equation}
where $\chi_{\text{tot}}$ is the total additional noise defined at the channel input, which can be calculated by $\chi_{\text{tot}}=\chi_{\text {line }}+\chi_{\text {het}}/T$, where $\chi_{\text {line }}=(1+T \varepsilon_\textup{A}) / T-1=1 / T-1+\varepsilon_\textup{A}$ represents the total channel-added noise defined at channel input, $\chi_{\mathrm{het}}=\left[1+(1-\eta)+2v_{\mathrm{el}}\right] / \eta$ represents the total detector-added noise defined at channel input, and~$T$ is the channel transmittance, in~which $\varepsilon_\textup{A}$ is the excess noise defined at channel input\cite{laudenbach2018continuous}, $v_{\mathrm{el}}$ is the electric noise of the detector, $\eta$ is the detection efficiency. As~for the calculation of $\chi_\textup{BE}$, it is as follows
\begin{equation}
 \renewcommand{\theequation}{B\arabic{equation}}
 \setcounter{equation}{3}
\chi_\textup{BE}=\sum_\textup{i=1}^{2} G\left(\frac{\lambda_\textup{i}-1}{2}\right)-\sum_\textup{i=3}^{5} G\left(\frac{\lambda_\textup{i}-1}{2}\right),
\end{equation}
where $G(x)=(x+1)\text {log }_{2} (x+1)-x\text {log }_{2} x$, $\lambda_{i}$ represent the symplectic eigenvalues of covariance matrices given by
\begin{equation}
 \renewcommand{\theequation}{B\arabic{equation}}
 \setcounter{equation}{4}
\begin{aligned}
\lambda_{1,2}^{2}&=\frac{1}{2}[A \pm \sqrt{A^{2}-4 B}],\\
\lambda_{3,4}^{2}&=\frac{1}{2}[C \pm \sqrt{C^{2}-4 D}],\\
\lambda_{5}&=1,    
\end{aligned}
\end{equation}
where
\begin{equation}
 \renewcommand{\theequation}{B\arabic{equation}}
 \setcounter{equation}{5}
\begin{aligned}
A &=V^{2}(1-2 T)+2 T+T^{2}\left(V+\chi_{\text {line }}\right)^{2} \\
B &=T^{2}\left(V\chi_{\text {line }}+1\right)^{2} \\
C &=\frac{1}{T^{2}\left(V +\chi_{\textup{tot}}\right)^{2}}[A\chi_{\textup{het}}^{2}+B+1\\
&+2\chi_{\textup{het}}\left(V\sqrt{B}+T(V+\chi_{\textup{line}})\right)+2 T(V^{2}-1)] \\
D &=\left(\frac{V+\sqrt{B} \chi_{\text{het}}}{T(V+\chi_{\text{tot }})}\right)^{2}
\end{aligned}
\end{equation}

In the case of finite-size effect, the channel transmittance $T$ would get a lower bound while the excess noise $\varepsilon_\textup{A}$ would reach a upper bound, written as

\begin{equation}
 \renewcommand{\theequation}{B\arabic{equation}}
 \setcounter{equation}{6}
\begin{aligned}
     T_\textup{min}&=\frac{\left(\sqrt{\eta T}-\Delta T\right)^{2}}{\eta},\\    \varepsilon_\textup{max}&=\varepsilon_\textup{A}+\frac{\Delta\sigma_\textup{0}^{2}}{\eta T},     
\end{aligned}
\end{equation}
where
\begin{equation}
 \renewcommand{\theequation}{B\arabic{equation}}
 \setcounter{equation}{7}
\begin{aligned}
\Delta T&=\zeta_\textup{$\delta_\textup{PE/2}$} 
\sqrt{\frac{\sigma^{2}}{m V_\textup{A}}}, \\
\Delta\sigma_\textup{0}^{2}&=\zeta_\textup{$\delta_\textup{PE/2}$} \frac{\sigma^{2}\sqrt{2}}{\sqrt{m}},
\end{aligned}
\end{equation}
where $\zeta_\textup{$\delta_\textup{PE/2}$}$ is the confidence coefficient, $\sigma^{2}=\eta T\varepsilon_\textup{A}+1+v_\textup{el}$ and $m=N-n$ is the number of the signal pulse for parameter estimation.
In addition, $\Delta(n)$ in Eq. B1 is related to the security of the privacy amplification and can be calculated as
\begin{equation}
 \renewcommand{\theequation}{B\arabic{equation}}
 \setcounter{equation}{8}
\Delta(n)= 7\sqrt{\frac{log_\textup{2}\left(2/\zeta\right)}{n}}+\frac{2}{n}log_\textup{2}\frac{1}{\zeta_\textup{PA}}.
\end{equation}

\begin{backmatter}
\bmsection{Funding} National Natural Science Foundation of China (62371060, 62001041, 62201012); State Key Laboratory of Information Photonics and Optical Communications (IPOC2022ZT09).

\bmsection{Disclosures} The authors declare no conflicts of interest.

\bmsection{Data Availability} Data underlying the results presented in this paper are not publicly available at this time but may be obtained from the authors upon reasonable request.

\end{backmatter}

\bibliography{sample}



\ifthenelse{\equal{\journalref}{aop}}{%
\section*{Author Biographies}
\begingroup
\setlength\intextsep{0pt}
\begin{minipage}[t][6.3cm][t]{1.0\textwidth} 
  \begin{wrapfigure}{L}{0.25\textwidth}
    \includegraphics[width=0.25\textwidth]{john_smith.eps}
  \end{wrapfigure}
  \noindent
  {\bfseries John Smith} received his BSc (Mathematics) in 2000 from The University of Maryland. His research interests include lasers and optics.
\end{minipage}
\begin{minipage}{1.0\textwidth}
  \begin{wrapfigure}{L}{0.25\textwidth}
    \includegraphics[width=0.25\textwidth]{alice_smith.eps}
  \end{wrapfigure}
  \noindent
  {\bfseries Alice Smith} also received her BSc (Mathematics) in 2000 from The University of Maryland. Her research interests also include lasers and optics.
\end{minipage}
\endgroup
}{}

\end{document}